\documentstyle[twocolumn,prb,aps,epsfig]{revtex}

\begin{document}
\draft
\title{Wave Chaotic Eigenfunctions in the Time-Reversal Symmetry-Breaking Crossover
Regime}
\author{Seok-Hwan Chung, Ali Gokirmak,\thanks{%
Present address: Cornell University, Department of Electrical Engineering,
Ithaca, NY} Dong-Ho Wu,\thanks{%
Present address: Naval Research Laboratory, Washington, DC} J. S. A.
Bridgewater, E. Ott, T. M. Antonsen and Steven M. Anlage}
\address{Department of Physics,\\
University of Maryland, \mbox{College Park, Maryland 20742-4111}}
\date{\today}
\maketitle

\begin{abstract}
We present experimental results on eigenfunctions of a wave chaotic system
in the continuous crossover regime between time-reversal symmetric and
time-reversal symmetry-broken states. The statistical properties of the
eigenfunctions of a two-dimensional microwave resonator are analyzed as a
function of an experimentally determined time-reversal symmetry breaking
parameter. We test four theories of one-point eigenfunction statistics and
introduce a new theory relating the one-point and two-point statistical
properties in the crossover regime. We also find a universal correlation
between the one-point and two-point statistical parameters for the crossover
eigenfunctions.
\end{abstract}





\narrowtext


Many complex quantum systems whose underlying classical behavior is chaotic
can be described by treating their Hamiltonian matrix elements as random
numbers which fluctuate around zero with a Gaussian distribution. There are
universal statistical properties of the eigenvalues and eigenfunctions of
these random matrices which depend only on the symmetries of the
Hamiltonian. For instance, random matrix theory has been shown to be
consistent with the statistical properties of nuclei,\cite{Haq} molecules,%
\cite{Zimm} and two-dimensional quantum dots.\cite
{Jalabert,F+E94,F+E96,Chang} In the case of two-dimensional quantum
mechanical systems with classically chaotic dynamics, there is a direct
relation between the statistics of measured conductance values through
quantum dots in the Coulomb blockade limit and the statistics of amplitudes
of chaotic electron waves in confined systems.\cite
{Jalabert,F+E94,F+E96,Chang}

When time-reversal symmetry is present, wave chaotic systems have
statistical properties described by a Gaussian Orthogonal Ensemble (GOE) of
random matrices.\cite{Leyvraz} As a magnetic field is applied, time-reversal
symmetry (TRS) is lost in these systems, and the statistical properties are
described by a Gaussian Unitary Ensemble (GUE) of random matrices. However,
it is found in the semiclassical regime that the evolution between these
types of symmetry is continuous and that a broad crossover of intermediate
statistics exists.\cite{BerryRob,Stone,Dupuis,Lenz} It has been proposed
that careful measurements of this crossover behavior provides a demanding
test of the random matrix hypothesis,\cite{Bohigas} and we perform such a
test in this paper. Prior experimental evidence for the existence of a
crossover regime comes from statistical properties of eigenvalue spacings,
which showed indications of a progression from GOE to GUE statistics as a
function of time-reversal symmetry breaking (TRSB) parameter.\cite
{BerryRob,French,So}

Here we address the evolution of eigenfunctions of semiclassical wave
chaotic systems from the TRS to the TRSB limits. A considerable theoretical
literature has developed proposing detailed descriptions of eigenvector
statistics in the crossover regime, although little experimental data is
available to test these theories. These theories only treat the evolution of
the one-point statistical property of eigenfunction distribution, P(%
\mbox{$\vert$}%
$\Psi $%
\mbox{$\vert$}%
$^2$), which quantifies the degree of probability density, 
\mbox{$\vert$}%
$\Psi $%
\mbox{$\vert$}%
$^2$, fluctuations in the eigenfunctions.\cite
{F+E94,Brickmann,Karol1,Sommers} No eigenfunction imaging experiment has
explicitly demonstrated the crossover of eigenfunction statistics from GOE
to GUE symmetry, to our knowledge. In addition, no work has addressed the
question of which of the theories of one-point eigenvector distribution
function best describes the crossover regime and no work has addressed the
crossover properties of the two-point correlation functions. Hence this work
forms an important testing ground for theories of wave chaotic systems and
quantum dots based on random matrix theory, supersymmetry, and semiclassical
techniques.

\begin{figure}[h]
\par
\begin{center}
\leavevmode
\epsfig{file=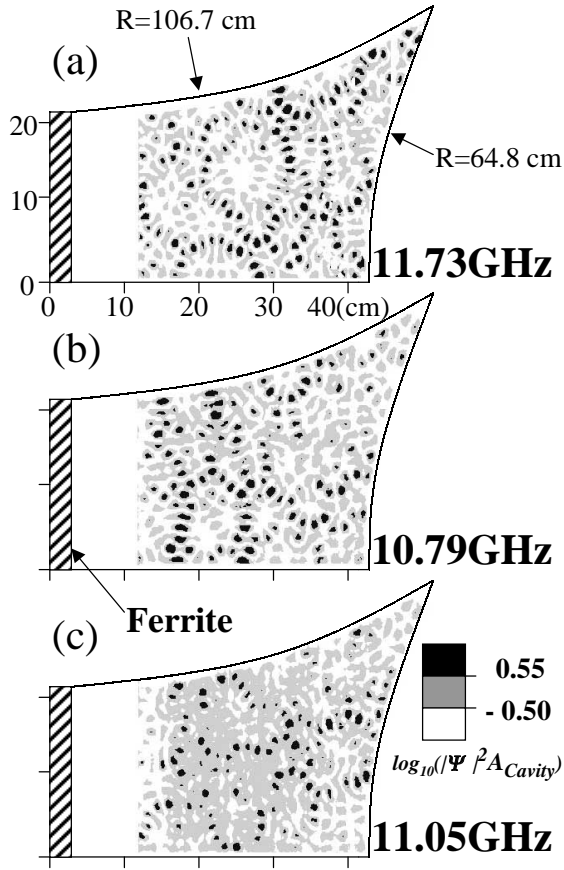,width=7.5cm,clip=,bbllx=232pt,bblly=59pt,bburx=551pt,bbury=559pt}
\end{center}
\caption{Experimentally determined two-dimensional probability amplitude
eigenfunctions plotted as log$_{10}$[ 
\mbox{$\vert$}%
$\Psi (x,y)$%
\mbox{$\vert$}%
$^2$$A_{cavity}$], derived from microwave resonator eigenmodes. The cavity
has a symmetry-reduced bow-tie shape of area $A_{cavity}$ and is loaded with
a magnetized ferrite on the left side. The images are taken at a) 11.73 GHz
(GOE), b) 10.79 GHz (crossover), and c) 11.05 GHz (GUE). The antenna are
located at (18.0, 15.5) and (45.5, 26.5) cm.}
\label{Fig1}
\end{figure}

The experimental arrangement used to create and measure the wave chaotic
eigenfunctions has been described previously.\cite{So,DongHo,AliRSI}
Briefly, a two-dimensional microwave cavity with walls defining a
non-integrable infinite square well potential is used to simulate the
solutions to the two-dimensional Schr\"{o}dinger equation in the
semi-classical limit.\cite{Stock} The cavity is a symmetry-reduced bow-tie
with dimensions shown in Fig. \ref{Fig1}(a).\cite{AliRSI} A magnetized
microwave ferrite incorporated into the cavity is used to break TRS. We
image the probability density 
\mbox{$\vert$}%
$\Psi (x,y)$%
\mbox{$\vert$}%
$^2$ by measuring the electric fields in the standing wave pattern of the
resonator and using the analogy between the Helmholtz and Schr\"{o}dinger
equations in two dimensions.\cite{DongHo} Previous results have established
that GOE\cite{Sridhar} and GUE statistical properties of both eigenvalues%
\cite{So,Stoff} and eigenfunctions\cite{DongHo} are seen in the limit of
zero and large non-reciprocal phase shift in the magnetized ferrite,\cite{So}
respectively.

We have found that the non-reciprocal property of the ferrite, hence the
degree of TRSB, is a function of frequency of the eigenmode in a relatively
narrow range of frequency.\cite{DongHo} This fortuitous property creates a
series of eigenmodes of similar energy spanning the GOE to GUE crossover
regime. Making use of this property, we identified a crossover of $\Delta _3$
spectral rigidity statistics from the GOE to GUE limits in earlier
experimental work.\cite{So} In this paper we systematically examine the
crossover eigenfunctions, and for the first time quantify the degree of TRSB
with an experimentally determined parameter.

Figure \ref{Fig1} shows three eigenmodes 
\mbox{$\vert$}%
$\Psi (x,y)$%
\mbox{$\vert$}%
$^2$ of the microwave resonator with different degrees of TRS, but of
similar energy. Figure \ref{Fig1}(a) shows a GOE-limit eigenmode, Fig. \ref
{Fig1}(c) shows a GUE-limit eigenmode, while Fig. \ref{Fig1}(b) shows a mode
of intermediate statistics. The probability amplitude is plotted as log$%
_{10} $(%
\mbox{$\vert$}%
$\Psi $%
\mbox{$\vert$}%
$^2A_{cavity}$) and presented in three shades to accentuate the differences
between the GOE\ and GUE characteristics of the eigenfunctions. Note that
the GOE mode is distinguished by the tall sharp fluctuations of 
\mbox{$\vert$}%
$\Psi $%
\mbox{$\vert$}%
$^2$ (dark areas), and the abundance of low 
\mbox{$\vert$}%
$\Psi $%
\mbox{$\vert$}%
$^2$ regions (white areas) between the spikes.\cite{Left} The GUE mode on
the other hand has smaller fluctuations of 
\mbox{$\vert$}%
$\Psi $%
\mbox{$\vert$}%
$^2$ and the peaks are more spread out, showing an abundance of intermediate
value 
\mbox{$\vert$}%
$\Psi $%
\mbox{$\vert$}%
$^2$ regions (gray areas).\cite{Left} These qualitative observations are
substantiated by more rigorous statistical analysis\cite{DongHo} presented
below. The crossover eigenfunction (Fig. \ref{Fig1}(b)) is a mixture of
co-existing regions showing GOE-like and GUE-like properties. Although we
see no obvious signs of scarring\cite{Heller,Antonsen} in any of the
eigenfunctions, it may make a contribution to the statistical measures of
crossover behavior presented below.\cite{Kaplan}

\begin{figure}[h]
\par
\begin{center}
\leavevmode
\epsfig{file=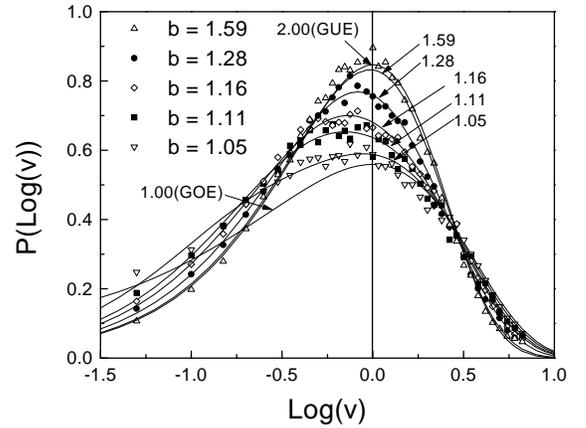,width=7.5cm,clip=,bbllx=30pt,bblly=27pt,bburx=700pt,bbury=533pt}
\end{center}
\caption{Probability amplitude distribution functions plotted as P(log$_{10}$%
($v$)) vs log$_{10}$($v$), where $v$ = 
\mbox{$\vert$}%
$\Psi $%
\mbox{$\vert$}%
$^2A_{cavity}$. Solid lines represent theoretical distribution functions in
the GOE, GUE, and crossover regimes described by the Zyczkowski and Lenz
theory P$_b $($v$). Also shown are averages of distribution functions for
data in the crossover regime.}
\label{Fig2}
\end{figure}

As mentioned above, the distribution of 
\mbox{$\vert$}%
$\Psi $%
\mbox{$\vert$}%
$^2$ values is a simple means of identifying the symmetry of the
corresponding Hamiltonian. In the GOE limit one finds the Porter-Thomas
distribution function of eigenfunction values,\cite{PT} P$_{GOE}$($v$) = (1/$%
\sqrt{2\pi v})e^{-v/2}$, where $v$ = 
\mbox{$\vert$}%
$\Psi $%
\mbox{$\vert$}%
$^2A_{cavity}$, and $A_{cavity}$ is the area of the cavity. In the GUE limit
the distribution function is simply P$_{GUE}$($v$) = $e^{-v}$.\cite
{PT,Prigodin} To enhance the subtle differences between these two
distribution functions, we plot P(log$_{10}$($v$)) versus log$_{10}$($v$) in
Fig. \ref{Fig2}.\cite{Brickmann} Note that P$_{GUE}$(log$_{10}$($v$)) peaks
at a higher value and falls off more quickly, while P$_{GOE}$(log$_{10}$($v$%
)) has a shorter and broader distribution. From this it is clear that GOE
eigenfunctions have more large (black in Fig. \ref{Fig1}) and small (white) 
\mbox{$\vert$}%
$\Psi $%
\mbox{$\vert$}%
$^2$ fluctuations, consistent with our qualitative observations of Fig. \ref
{Fig1}. Also shown in Fig. \ref{Fig2} are distribution functions of
eigenmodes in the crossover regime. Each data set is an average of 5
eigenfunctions of similar degree of TRSB (as discussed below). There is a
smooth variation of the distribution of eigenfunction fluctuations as TRS is
broken.\cite{Variance} The exact form of this variation is not predicted by
theory.

We have found four theories for the GOE $\rightarrow $ GUE crossover
probability density distribution function P$_{crossover}$($v$); that due to
Brickmann, {\it et al}. (P$_\beta $($v$), $\beta \in (1,2)$),\cite{Brickmann}
Zyczkowski and Lenz (P$_b$($v$), $b\in (1,2)$),\cite{Karol1} Sommers and
Iida (P$_\varepsilon $($v$), $\varepsilon \in (0,\infty )$),\cite{Sommers}
and Fal'ko and Efetov (P$_X$($v$), $X\in (0,\infty )$).\cite{F+E94} The
subscript denotes the symmetry-breaking parameter used by the authors. In
each case, the lower limit of the TRSB parameter ($\beta $, $b$, $%
\varepsilon $, $X$) corresponds to GOE statistics, and the upper limit to
GUE statistics. All of the theories agree in these two limiting cases, but
they disagree for intermediate parameter values in the crossover regime.\cite
{Sommers} For instance P$_\beta $($v$) $=\chi _\beta ^2\left( v\right) $, a
generalized chi-square distribution suggested by Brickmann, {\it et al}.,%
\cite{Brickmann} shows a maximum of the distribution function P$_\beta $($v$%
) at $v$ = 1 for all values of $\beta $, whereas the other theories show a
peak in P$_{b,\varepsilon ,X}$($v$) which occurs for $v$ $<$ 1 in the
crossover regime (see e.g. the solid lines in Fig. \ref{Fig2} for P$_b$($v$%
)).

It should be noted that none of the theories for the crossover behavior is
precisely aplicable to this experiment. The theory of Brickmann, {\it et al}%
. proposed an interpolation of the chi-square distribution with no
microscopic justification. Zyczkowski and Lenz performed a more detailed
calculation in which two chi-square distributions were convolved together
because it was assumed that the real and imaginary parts of the random
matrix elements were independently fluctuating. Sommers and Iida addressed
the issue of properly treating the fluctuations of the matrix elements, but
had to perform an energy averaging to arrive at the final distribution
function. The work of Fal'ko and Efetov uses a model appropriate for
disordered systems and was not explicitly derived for ballistic billiard
systems such as ours. (However the results here suggest that the theory is
also applicable to such systems.) Despite the fact that none of these
theories exactly corresponds to the experimental situation, they are
remarkably accurate in their description of the data.

The probability distribution functions P($v$) for 64 experimental eigenmodes
have been fit to the four theories of crossover statistics. It is found that
the Zyczkowski and Lenz theory produces the best fits for these
eigenfunctions, although the difference in fit quality for the other
theories is not statistically significant. Figure \ref{Fig2} shows averaged
distribution functions for groups of 5 eigenfunctions with similar values of
the Zyczkowski and Lenz crossover parameter $b$, as well as best fits to the
averaged P$_b$($v$). Notice that significant changes in P($v$) already occur
for a small deviation of $b$ from 1, consistent with Fal'ko and Efetov's
prediction that a small amount of magnetic flux quickly moves the system
away from GOE statistics.\cite{F+E94} Also note that the peak of the
experimental distribution function occurs at $v$ $<$ 1 for crossover
eigenfunctions, demonstrating that the generalized $\chi _\beta ^2\left(
v\right) $ distribution is not a correct description of the crossover data.

These theories, as well as random matrix theory,\cite{Bohigas} do not
predict how the crossover parameters evolves with non-reciprocal phase shift
in the ferrite, and do not even relate their crossover parameters to those
of other theories. Therefore we proceed empirically and define a simple
experimental measure of the degree of TRSB. We have noticed a strong
asymmetry of the forward (S$_{12}(f)$) and reverse (S$_{21}(f)$) complex
transmission coefficients of the microwave cavity for TRSB eigenmodes,\cite
{AliRSI} and correlated this asymmetry with the degree of TRSB derived from
statistical analysis of the associated eigenfunctions. We define the
experimental time-reversal asymmetry parameter $A=\int \left( \left|
|S_{12}|-|S_{21}|\right| \right) df/\int \left( |S_{12}|+|S_{21}|\right) df$%
, where the integrals are carried out over one resonant peak (between
neighboring minima in $|S_{12}(f)|$) in the frequency domain. This parameter
is easily evaluated, does not require an image of the eigenmode,\cite
{Footnote} and (as shown below) can be considered a measure of the
``magnetic flux'' through the cavity causing TRSB.

\begin{figure}[h]
\par
\begin{center}
\leavevmode
\epsfig{file=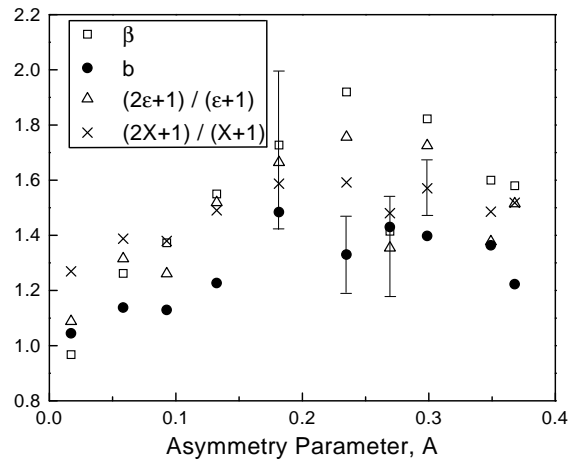,width=7.5cm,clip=,bbllx=98pt,bblly=48pt,bburx=699pt,bbury=532pt}
\end{center}
\caption{Plot of crossover parameters $\beta $, $b$, $(2\varepsilon +1)/(%
\varepsilon +1)$, and $(2X+1)/(X+1) $ versus the experimentally determined TRSB
parameter $A$. The GOE limit is in the lower left, while the GUE limit is in
the upper right of the figure. The error bars show the typical standard
deviation of the data points.}
\label{Fig3}
\end{figure}

We find that the one-point correlation function parameters ($\beta $, $b$, $%
\varepsilon $, $X$) are strongly correlated with our experimental measure of
TRSB, $A$. Figure \ref{Fig3} shows the four distribution function crossover
parameters $\beta $, $b$, ($2\varepsilon +1)/(\varepsilon +1)$, $(2X+1)/(X+1)
$ plotted vs. $A$, for groups of similar eigenfunctions. Note that the
parameters $\varepsilon $ and $X$ have been mapped to the interval (1,2) in
an {\it ad hoc} way, simply for comparison with the other statistical
parameters. We see that all four statistical parameters describe a smooth
and universal transition from GOE to GUE statistics as the asymmetry of the
transmission characteristic, $A$, increases.\cite{FiniteData} Note that
there is initially a linear increase of the crossover parameters with A,
suggesting that they all give a measure of the ``magnetic flux'' producing
the GOE $\rightarrow $ GUE crossover. The reduction of the parameter values
beyond $A$ = 0.25 is likely due to difficulties in properly calculating A
for the highly distorted microwave resonance curves $S_{12}(f)$ encountered
in the GUE limit.

One can also employ two-point statistical correlation functions to quantify
the crossover from TRS to TRSB behavior of the eigenfunctions. It has been
shown that $C(kr)=\langle |\Psi (0)\Psi (r)|^2\rangle \sim $ $1+cJ_0^2(kr)$,%
\cite{Berry} where $k$ is the wavenumber of the eigenmode, $J_0$ is the
zero-order Bessel function, and $c$ = 2 for GOE and $c$ = 1 for GUE
eigenfunctions.\cite{DongHo,Prigodin} Eigenfunctions with a smaller value of 
$c$ show a more ``smeared out'' appearance (Fig. \ref{Fig1}(c)). We find
that the value of $c$ changes smoothly between 2 and 1 for the experimental
crossover eigenfunctions.

\begin{figure}[h]
\par
\begin{center}
\leavevmode
\epsfig{file=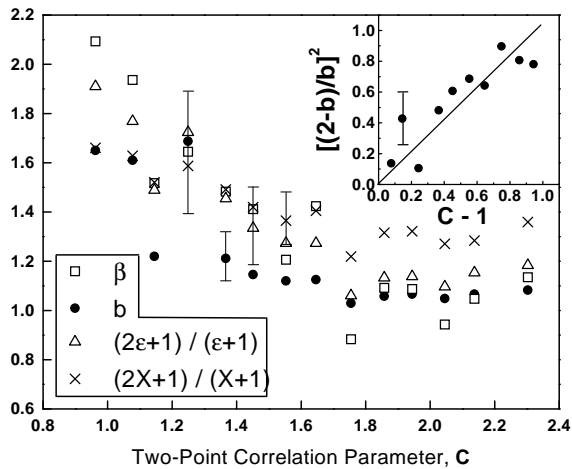,width=7.5cm,clip=}
\end{center}
\caption{Plot of crossover parameters $\beta $, $b$, $(2 \varepsilon +1)/(%
\varepsilon +1)$, and $(2X+1)/(X+1)$ versus the two-point correlation function
parameter $c$. The GUE limit is in the upper left, while the GOE limit is in
the lower right of the figure. The error bars show the typical standard
deviation of the data points. The inset shows a correlation between the
one-point statistical parameter $b$, and the two-point parameter $c$, in the GOE
to GUE crossover regime.}
\label{Fig4}
\end{figure}

To investigate the self-consistency of these results, we have examined the
relationship between the one-point correlation function parameters ($\beta $%
, $b$, $(2\varepsilon +1)/(\varepsilon +1)$, $(2X+1)/(X+1)$) and the two-point
parameter $c$ (determined by fitting $C(kr)$), as shown in Fig. \ref{Fig4}.
Again we notice that there is a universal behavior of the one-point
crossover parameters in their dependence on $c$.\cite{FiniteData} This
demonstrates that the crossover behavior is shared by both single point and
two-point statistical properties of the eigenfunctions, in agreement with
the prediction of the supersymmetric nonlinear $\sigma $ model.\cite
{F+E94,F+E96}

To develop a relationship between one-point and two-point eigenfunction
properties, consider a superposition of plane waves,\cite{Berry} $\Psi \sim
\sum a_{{\bf k}}e^{i{\bf k}\cdot {\bf x}}$ , with fixed wavenumber, $k=|{\bf %
k}|$, but with random directions and amplitudes. Writing $a_{{\bf k}}=|a_{%
{\bf k}}|e^{i\theta _k}$, GOE corresponds to 
\mbox{$\vert$}%
$a_{{\bf k}}|=|a_{-{\bf k}}|$ and $\Delta _{{\bf k}}\equiv \theta _{{\bf k}%
}+\theta _{-{\bf k}}$ zero (to make $\Psi $ real), while GUE corresponds to $%
\Delta _{{\bf k}}$ random in $\left[ 0,2\pi \right] $. Allowing general $%
\Delta _{{\bf k}}$ in the crossover regime leads to the two-point result for 
$C(kr)$ given above with $c=1+|\sum |a_{{\bf k}}||a_{-{\bf k}}|\exp (i\Delta
_k)/\sum |a_{{\bf k}}|^2|^2$, and to the one point result of Zyczkowski and
Lenz, P$_b$(v), with $c=1+$ $\left( \frac{2-b}b\right) ^2$. To test this
prediction, we plot $c-1$ versus $\left( \frac{2-b}b\right) ^2$ as an inset
to Fig. \ref{Fig4}. The data points fit to a straight line of slope close to
1, demonstrating remarkable agreement between the one-point and two-point
statistical parameters in the crossover regime.

To summarize, we have established both GOE (TRS) and GUE (TRSB) properties
of wave chaotic eigenfunctions and shown that there is a continuum of
eigenfunctions with intermediate statistics between the two time-reversal
symmetry states. We have found that three of the four theories of one-point
eigenfunction statistics P($v$) in the crossover regime adequately describe
our data. We have introduced a simple experimental quantity which measures
the degree of TRSB, and shown that the statistical properties of the
eigenfunctions evolve smoothly with an increase of this parameter. Finally
we demonstrate that the one-point and two-point correlation functions
describe the smooth crossover in a consistent and universal manner, and that a 
single TRSB parameter must describe the crossover behavior.

We wish to acknowledge assistance from Paul So and Karol Zyczkowski. This
work has been supported by an NSF NYI grant \# DMR-9258183, Mitre Corp., and
by the Maryland Center for Superconductivity Research.



\begin{references}
\bibitem{Left}  The leftmost 4 inches of the cavity is not imaged due to the
interference of the scanning magnet with the ferrite. See reference \cite
{AliRSI}.
\bibitem{Footnote}  We have also evaluated a more comprehensive asymmetry
parameter B = $\int \left( \left| |\Psi _{12}|^2-|\Psi _{21}|^2\right|
\right) dA_{cavity}/\int \left( |\Psi _{12}|^2+|\Psi _{21}|^2\right)
dA_{cavity}$, which makes use of a pair of entire eigenfunction images. We
find that the quantities A and B are strongly correlated.
\bibitem{FiniteData}  Note that the failure of the data points to reach the
full GOE and GUE limits in Figs. \ref{Fig3}, \ref{Fig4} is due to the finite
experimental resolution for the probability amplitude: $0.05\leq v\leq 10$.%
\cite{DongHo}
\end{references}
\end{document}